

\magnification=\magstep1

\footline = {\ifnum \pageno > 1 \hfil {\rm \folio} \hfil
             \else              \hfil \fi}

\font\title=cmbx10
\font\capsone=cmcsc10 at 12pt 
\font\scaps=cmcsc10           

\def\titlea#1#2{\vskip 20pt \noindent{\title #1 #2}\vskip 10pt \noindent}

\rightline{CERN-TH.7470/94}
\rightline{hep-th/9410239}
\vskip 80pt

\centerline{\capsone AUXILIARY FIELDS FOR SUPER YANG-MILLS}
\vskip 5pt
\centerline{\capsone FROM DIVISION ALGEBRAS}
\vskip 50pt
\centerline{\scaps
Jonathan M.~Evans\footnote{*}{\rm Supported by a fellowship
from the EU Human Capital and Mobility programme.}
}
\vskip 10pt
\centerline{\sl
Theoretical Physics Division}
\centerline{\sl CERN}
\centerline{\sl CH-1211 Gen\`eve 23}
\centerline{\sl Switzerland}
\vskip 5pt
\centerline{ {\sl e-mail}: evansjm@surya11.cern.ch}
\vskip 50pt
\centerline{\capsone ABSTRACT}
\vskip 10pt

\noindent
\phantom{xxxx}Division algebras are used to explain the existence
and symmetries of various
\hfil \break
\phantom{xxxx}sets of auxiliary fields for super Yang-Mills
in dimensions $d=3,4,6,10$.

\vskip 50pt

\centerline{\sl To appear in the proceedings of}
\centerline{\sl G\"ursey Memorial Conference I: Strings and Symmetries}
\centerline{\sl Bo\v gazi\c ci University, Istanbul, Turkey}
\centerline{\sl 6th-10th June 1994}
\vfil

\line{CERN-TH.7470/94 \hfil}
\line{October 1994 \hfil}
\eject

\font\bbb=msbm10
\def\R{ {\hbox{\bbb R}} }
\def\C{ {\hbox{\bbb C}} }
\def\H{ {\hbox{\bbb H}} }
\def\O{ {\hbox{\bbb O}} }
\def\K{ {\hbox{\bbb K}} }
\def\a{\alpha}
\def\ad{{\dot \alpha}}
\def\eps{\epsilon}
\def\dag{\dagger}
\def\half{ {\textstyle{1\over2}} }
\def\del{\partial}

\titlea{1.}{Introduction}The
simplest supersymmetric Yang-Mills theories are those in which the
physical degrees of freedom are described by a vector gauge field
$A_\mu$ $(\mu = 0, \ldots , d{-}1)$ and a spinor field $\psi$,
both defined on $d$-dimensional Minkowski space and
taking values in the Lie algebra of some gauge group.
A necessary condition for supersymmetry is that the
number of physical bosons and fermions should be equal,
which leads to the possibilities
$d = 3, 4, 6, 10$ with the spinor being
Majorana, Majorana or Weyl, Weyl, Majorana and Weyl respectively.
To show that this equality of bosons and fermions
is sufficient for supersymmetry as well as necessary,
one can take the detailed expressions for the
supersymmetry transformations---which are fixed up to irrelevant constants
by gauge-invariance and dimensional considerations---and check that
they obey the standard algebra up to terms involving
field equations
(this is in fact equivalent to requiring invariance
of the natural action in which
the spinor is minimally coupled to the gauge field).
The condition for the supersymmetry algebra
to close {\it on-shell\/} in this way is a
certain gamma-matrix identity which is indeed satisfied precisely for the
values of $d$ and the types of spinor listed above [1].

It is always desirable in a supersymmetric theory to try to promote the
on-shell symmetry algebra to one which holds
{\it off-shell\/}, that is, without the use of field equations.
For $d=3$ super Yang-Mills the superalgebra closes automatically;
for $d = 4, 6$ one can introduce auxiliary fields to
close the superalgebra in a Lorentz-covariant
way [2,3]; but for $d=10$ the best which can be done with a finite set of
auxiliary fields is to partially close the superalgebra,
breaking the manifest $d=10$ Lorentz symmetry to
some subgroup in the process.
An interesting new perspective on these matters was
provided recently in [4] with the introduction of more general
fermionic transformations which include the conventional
supersymmetry transformations of $d=10$ super Yang-Mills as special
cases. It was then shown in [5] that all previously known sets of
auxiliary fields for the $d=10$ theory could be recovered
within this framework.

Our aim here is to show how the possible auxiliary fields for each of
the allowed super Yang-Mills theories can be understood using the language of
division algebras [6-8].
Connections between super Yang-Mills and the division algebras have been
established in the past by interpreting the gamma-matrix identity
necessary for on-shell supersymmetry from a number of closely related
points of view using division algebra-valued spinors [9,10], Jordan
algebras [10,11] and trialities [12].
The role of division algebras in understanding off-shell
supersymmetry was emphasized first in [7] and then in [4] where
octonions were used to find a new
set of auxiliary fields in $d=10$. We shall indicate here how the
solutions given in [4,5] for $d=10$, together with analogous solutions for the
lower-dimensional theories, can all be understood using this language;
we shall also explain from this point of view
the residual symmetries of these various sets of auxiliary fields.

\titlea{2.}{Division Algebra Notation}We
denote by $\K_n$ with $n=1,2,4,8$ the division algebras
$\R$, $\C$, $\H$, $\O$ of real numbers, complex numbers, quaternions
and octonions respectively; for background see papers such as [6-12].
We denote by $e_a$ $(a=1, \ldots , n)$ an orthonormal basis for $\K_n$
with $e_n = 1$ and all other basis elements pure-imaginary. Bars will
denote conjugation and daggers will denote hermitian conjugation.

The basic idea is to define the action of certain spin representations
of SO($n{+}1,1)$ on two-component objects
with entries in $\K_n$ [7,8].
The cases $n = 1, 2, 4, 8$ obviously correspond to
$d= 3, 4, 6, 10$ and the spinor appearing in each of the allowed
super Yang-Mills theories in these dimensions can then be written as
an object $\psi^\a$ ($\a=1,2$)
or as its conjugate $\bar \psi^{\ad}$ ($\ad=1,2$).
There are also dual spinor representations acting on objects which we
denote by $\chi_\a$ and $\bar \chi_\ad$.
When spinor indices are suppressed we will regard $\psi$, $\bar
\psi$ as row vectors and $\chi$, $\bar \chi$ as column vectors.
Spinor indices are never raised or lowered and the usual
Lorentz-invariant inner-product is given by the real
expression $\psi \chi + \chi^\dag \psi^\dag$ =
$\psi^\a \chi_\a + \bar \chi_\ad \bar \psi^\ad$.

The gamma-matrices needed to construct the spin representations
are invariant tensors
$(\Gamma_\mu)_{\a \ad}$ and $(\Gamma^\mu)^{\ad \a}$.
We define these in a particular basis in which their components
are equal and are
given by the hermitian matrices:
$$
\pmatrix{1 & 0 \cr 0 & 1 \cr} \, {\rm for} \, \, \mu = 0 ; \quad
\pmatrix{1 & 0 \cr 0 & \! \!-1 \cr} \, {\rm for} \, \, \mu = n{+}1 ; \quad
\pmatrix{0 & \! e_a \cr \bar e_a & \! 0 \cr} \, {\rm for} \, \, \mu = a = 1,
\ldots, n .
$$
The order of indices on gamma-matrices indicates whether
they should multiply spinors from the left or from the right with the
convention that only adjacent upper and lower indices of the same
type can be contracted.
For example, the matrices
$(\Gamma_\mu)_{\a \ad} $ act by
right-multiplication on $\psi^\a$ and left-multiplication on
$\bar \psi^{\ad}$.

It will be important for us to understand how certain subgroups
of SO($n{+}1,1$) appear in this formalism.
With the basis introduced above, the two $\K_n$-valued components
of any spinor clearly carry
irreducible representations of the light-cone subgroup SO(1,1)$\times$SO($n$).
In fact these two
copies of $\K_n$ carry the inequivalent spin representations of SO($n$)
for $n = 2, 4, 8$.
If we restrict further to the subgroup
SO($n{-}1$) which fixes the direction $\mu = n$ then these spin
representations become equivalent.
The resulting representation is realized on $\K_n$ in two particularly
simple ways: the relevant gamma-matrices or invariant tensors are just
the imaginary basis elements $e_i$ ($i = 1, \ldots , n{-}1$) and
these multiply components of type $\psi^\a$ from the right and
components of type $\chi_\a$ from the left.

Notice that the elements of unit modulus in $\C$ and $\H$ form groups
U(1) and SU(2) respectively. They act naturally by multiplication on
complex and quaternionic spinors so as to commute with Lorentz
transformations; we shall therefore refer to these operations on
spinors as internal transformations.
The generators of such transformations are unit imaginary elements $e_i$
multiplying components of type
$\psi^\a$ from the left and components of type $\chi_\a$ from the
right (compare with the last paragraph).
The internal groups U(1) and SU(2) appear as
symmetries of the super Yang-Mills theories in $d=4$ and $d=6$;
no such symmetries arise in $d=3$ or $d=10$.

Finally, it can be shown that the algebras $\H$ and $\O$
have non-trivial continuous automorphism groups
SO(3) and G${}_2$ respectively [6,8].
The action of these automorphisms on spinor components can always be
reproduced by combinations of Lorentz
and internal transformations.
The special feature of
the octonionic case is that the automorphism group is contained entirely
within the Lorentz group---in fact within the SO(7) subgroup mentioned
above.

\titlea{3.}{Generalized and Conventional Supersymmetry}The
generalized supersymmetry transformations proposed in [4] involve
real bosonic auxiliary fields $G_i$ $(i =1 , \ldots , d{-}3)$ which
balance the off-shell bosonic and fermionic degrees of freedom.
They can be written
$$\eqalign{
\delta A_\mu & =
\eps (\Gamma_\mu \psi^\dag) + (\psi \Gamma_\mu) \eps^\dag \, ,\cr
\delta \psi & =
\half (\eps \Gamma_\nu) \Gamma_\mu F^{\mu \nu}
+ G_i v_i \, ,\cr
\delta G_i & =
- v_i (\Gamma^\mu D_\mu \psi^\dag)
- ( D_\mu \psi \Gamma^\mu) v^\dag_i
\, , \cr
}\eqno(1)$$
where $D_\mu = \del_\mu + A_\mu$ is the usual covariant derivative and
$F_{\mu \nu} = [ D_\mu , D_\nu ]$ is the field strength.
(We have defined $\delta$ so as to remove some factors of $i$ compared
to [4,5] and we have normalized spinors so as to remove some
factors of $\half$.)
The parameters of the transformations are commuting spinors $\eps^\a$
and $v^\a_i$ which must satisfy the additional relations
$$\eqalign{
\eps (\Gamma_\mu v_i^\dag) + (v_i\Gamma_\mu) \eps^\dag
& = 0 \, , \cr
v_i (\Gamma_\mu v_j^\dag ) + (v_j \Gamma_\mu ) v_i^\dag & =
\delta_{ij}
(\, \eps (\Gamma_\mu \eps^\dag ) + ( \eps \Gamma_\mu ) \eps^\dag \, )
\, . \cr
}\eqno(2)$$
These ensure that the standard supersymmetry algebra still
holds up to field equations despite the introduction of extra
parameters.

To recover conventional supersymmetry
transformations from those written in (1) above one must
solve the equations (2) with $v_i$ depending linearly on
$\eps$ and with the value of $\eps$
restricted to some subspace if necessary.
The subset of conventional supersymmetry transformations
obtained in this way will automatically obey a closed algebra.
However, such a solution will, in general, break
the full Lorentz invariance of equations (1) and (2)
down to a subgroup determined both by the subspace to which $\eps$
is confined and by the precise definitions of the quantities $v_i$.
These points are explained in more detail in [5].

\titlea{4.}{Solutions and their Symmetries}Solutions to (2) can be
written
very simply in division algebra
notation. We consider first the possibility
$$
v_i = e_i \eps
\eqno (3)
$$
with the spinor $\eps$ confined to some subspace.
The transformations (1) can now be re-expressed in a more
compact way by combining the $n{-}1$ auxiliary fields into the
pure-imaginary object $G= G_i e_i$ with the result
$$\eqalign{
\delta A_\mu & =
\eps (\Gamma_\mu \psi^\dag) + (\psi \Gamma_\mu) \eps^\dag \, , \cr
\delta \psi & = \half (\eps \Gamma_\nu) \Gamma_\mu F^{\mu \nu} + G
\eps \, , \cr
\delta G & =
\eps (\Gamma^\mu D_\mu \psi^\dag) -
( D_\mu \psi \Gamma^\mu) \eps^\dag \, .
\cr
}\eqno(4)$$
Before discussing the individual solutions it will be best to make
some general remarks concerning the possible symmetries of these equations.

In the associative cases the equations (4) are invariant under any
Lorentz transformation if the auxiliary fields behave as scalars
(and provided that the transformation respects whatever
restriction is placed on $\eps$ of course).
This can be checked by calculating explicitly the transformation of
the expression given for $\delta G$
and by noting also that if $G$ is inert then the term $G \eps$
transforms in the same way that $\eps$ does; all other terms are
covariant by construction.
In the octonionic case, however, this statement must be qualified:
we find that a Lorentz transformation is a
symmetry of (4) with $G$ inert only if it is
constructed from gamma-matrices whose entries have vanishing
associators with the components of $\eps$.
Aside from such Lorentz transformations, the
equations (4) are also invariant under the internal transformations
which arise in the complex and quaternionic cases in the manner
discussed earlier.
Lastly, the equations (4) are invariant under
any automorphism of $\K_n$ (provided once again that it respects the
restriction on $\eps$) and then the auxiliary
fields will transform in some non-trivial way.
{}From the remarks made previously we know
that automorphisms give genuinely new symmetries
only in the octonionic case.
We are now in a position to explain the symmetries which appear in
each of our solutions.

It is easy to show that in the associative cases (3) provides a
solution of (2) for any spinor $\eps$ and the
formulas (4) then reproduce the standard, Lorentz-covariant,
off-shell supersymmetry transformations for $d = 4$ [2] and $d = 6$ [3,7]
super Yang-Mills with scalar auxiliary fields.
In $d=4$ the off-shell supersymmetries are also invariant under
the U(1) internal group with the auxiliary field being inert.
In $d=6$ they are invariant under
the SU(2) internal group with the auxiliary
fields transforming as a triplet.

The lack of associativity of the octonions means that in $d = 10$ the formula
(3) is no longer a solution of (2) for all $\eps$.
However, there are at least two
interesting ways of restricting $\eps$ in (3) so as to obtain
closed subalgebras of supersymmetry transformations of type (4).
Both solutions require for their verification several lines of octonionic
manipulations which we shall omit.

The first case in which (3) provides a solution of (2)
is where one of the components of $\eps$ is restricted to be
real, giving a closed algebra of nine supersymmetries [4].
To preserve the form of $\eps$ under a Lorentz transformation we
must confine attention to a subgroup SO(1,1)$\times$SO(7).
The generator of SO(1,1) is real and so never gives rise to an
associator.
But the additional transformations are symmetries of (4) only if they
lie in the subgroup of automorphisms G${}_2$ within SO(7), with the auxiliary
fields transforming in its
seven-dimensional representation. In this way we find exactly the
residual invariance SO(1,1)$\times$G${}_2$ and the pattern of
representations given in section 3 of [5].

The second case in which (3) provides a solution of (2) is
where both components of $\eps$ lie
in some copy of the complex numbers $\C$ within $\O$, yielding
a closed algebra of four supersymmetries.
This solution is clearly invariant under a subgroup SO(3,1) generated
by combining gamma-matrices with entries in this same copy of $\C$,
since all the relevant associators then vanish.
The other pairs of gamma-matrices
which lead to vanishing associators give
$d=10$ Lorentz generators which coincide when acting on $\eps$
and which, when exponentiated, correspond simply to multiplication by an
arbitrary phase within our chosen copy of $\C$.
Lastly, the solution is invariant under the subgroup of
G${}_2$ which sends our particular copy of $\C$ to
itself, and it is
well-known that the subgroup of automorphisms
which fix a given imaginary
element is SU(3) [6].
The residual
invariance is therefore
SO(3,1)$\times$U(1)$\times$SU(3) and closer examination of the details of the
representations gives complete agreement with those found
in section 4 of [5].

So far we have discussed solutions of (2) of type (3); it is
natural to ask whether there exist similar
solutions of the form
$$
v_i = \eps e_i
\eqno (5) $$
with $\eps$ restricted to some subspace.
The transformations (1) can once again be re-written, after a little work,
in terms of $G= G_i e_i$ and the result is
$$\eqalign{
\delta A_\mu & =
\eps (\Gamma_\mu \psi^\dag) + (\psi \Gamma_\mu) \eps^\dag \, ,
\cr
\delta \psi & = \half (\eps \Gamma_\nu) \Gamma_\mu F^{\mu \nu}
+ \eps G \, , \cr
\delta G & =
(\overline{D_\mu \psi \Gamma^\mu}) {\bar \eps}^\dag
- \bar \eps (\overline{\Gamma^\mu D_\mu \psi^\dag})
\, . \cr
}\eqno(6)$$
In $d=4$ the formulas (5) and (6) coincide with (3) and (4)
because the complex numbers are commutative; but for $d=6$ and $d=10$
we obtain new possibilities.

In the non-commutative cases (5) fails to satisfy (2) for
general $\eps$, but it does provide a solution if we restrict
$\eps$ to having only one non-zero component.
This restriction clearly breaks the Lorentz
group at least to the light-cone subgroup
SO(1,1)$\times$SO($n$) and in fact the compact part of the surviving
symmetry is SO($n{-}1$). This can be seen from the way in which
the pure-imaginary elements $e_i$ appear in (6) through $G$, because
these quantities act as invariant tensors for the subgroup SO($n{-}1$) when
multiplying $\eps$ from the right, as we have already mentioned.
Thus in $d=6$ we obtain a closed algebra of four supersymmetries of
type (6) with
residual Lorentz invariance SO(1,1)$\times$SO(3) and
auxiliary fields transforming as a three-dimensional vector.
There is also an SU(2) internal symmetry under which the auxiliary
fields are inert. In $d =10$ we find a closed algebra
of eight supersymmetries of type (6) with residual invariance
SO(1,1)$\times$SO(7) and auxiliary fields
transforming as a seven-dimensional vector.
This is just the solution presented in section 2 of [5].

\titlea{5.}{Comments}We have seen that division algebras provide an
elegant language in which to
understand the occurrence and symmetries of bosonic
auxiliary fields for super Yang-Mills, with the lack of a covariant
set in $d=10$ being traced directly to the non-associativity of the
octonions [4].
It is natural to wonder to what extent the solutions we have discussed
here are exhaustive.
It would also be interesting to
investigate whether division algebras
could be used to understand fermionic auxiliary
fields (see the papers cited in [5]) in a similar fashion.
\vskip 20pt

\centerline{\capsone ACKNOWLEDGEMENTS}
\vskip 10pt
\noindent
I thank Nathan Berkovits for interesting discussions and correspondence.
I am also grateful to the organizers of the G\" ursey Memorial
Conference I: Strings and Symmetries for giving me the opportunity
to speak at this stimulating meeting.
\vskip 20pt

\centerline{\capsone REFERENCES}
\vskip 10pt

\item{[1]} L. Brink, J. Scherk, J.H. Schwarz: Nucl. Phys. B {\bf 121}
77 (1977);
\hfil \break
M.B. Green, J.H. Schwarz, E. Witten: {\it Superstring Theory\/},
C.U.P. (1987).
\item{[2]} M.F. Sohnius: Phys. Rep. {\bf 128} 39 (1985).
\item{[3]} P. Howe, G. Sierra, P.K. Townsend: Nucl. Phys. B {\bf
221} 331 (1983).
\item{[4]} N. Berkovits: Phys. Lett. B {\bf 318} 104 (1993).
\item{[5]} J.M. Evans: Phys. Lett. B {\bf 334} 105 (1994).
\item{[6]} M. G\"unaydin, F. G\"ursey: J. Math. Phys. {\bf 14} 1651 (1973).
\item{[7]} T. Kugo, P.K. Townsend: Nucl. Phys. B {\bf 221} 357 (1983).
\item{[8]} A. Sudbery: J. Phys. A {\bf 17} 939 (1984);
\hfil \break
K.-W. Chung, A. Sudbery: Phys. Lett. B {\bf 198} 161 (1987).
\item{[9]} C.A. Manogue, A. Sudbery: Phys. Rev. D {\bf 40} 4073 (1989);
\hfil \break
D.B. Fairlie, C.A. Manogue: Phys. Rev. D {\bf 36} 475 (1987).
\item{[10]} R. Foot, G.C. Joshi: Mod. Phys. Lett. A {\bf 3} 999 (1988).
\item{[11]} G. Sierra: Class. Quantum Grav. {\bf 4} 227 (1987).
\item{[12]} J.M. Evans: Nucl. Phys. B {\bf 298} 92 (1988).

\bye